\documentclass[%
 reprint,
superscriptaddress,
showpacs,
preprintnumbers,
amsmath,
amssymb,
aps,
floatfix,
]{revtex4-1}
\RequirePackage[detect-mode,detect-weight]{siunitx}
\usepackage{graphicx}
\usepackage{dcolumn}
\usepackage{bm}
\usepackage{natbib}
\usepackage{xcolor}
\usepackage{wasysym}
\newcommand{\hmm}{\ensuremath{\text{H}^-\ }}
\newcommand{\hz}{\ensuremath{\text{H}^0\ }}
\begin{document}
\bibliographystyle{unsrt}
\title{Demonstration of a laserwire emittance scanner for the CERN LINAC4 $\text{H}^-$ Beam}
\author{T. Hofmann}
\email{thomas.hofmann@cern.ch}
\affiliation{Beam Instrumentation Group, CERN, CH-1211 Geneva 23, Switzerland}
\author{K.O. Kruchinin}
\author{A. Bosco}
\author{S.M. Gibson}
\affiliation{John Adams Institute at Royal Holloway, University of London, Egham, TW20 0EX, United Kingdom}
\author{F. Roncarolo}
\affiliation{Beam Instrumentation Group, CERN, CH-1211 Geneva 23, Switzerland}
\author{G.~Boorman}
\affiliation{John Adams Institute at Royal Holloway, University of London, Egham, TW20 0EX, United Kingdom}
\author{U. Raich}
\author{E. Bravin}
\affiliation{Beam Instrumentation Group, CERN, CH-1211 Geneva 23, Switzerland}
\author{J.K. Pozimski}
\affiliation{Imperial College London, SW7 2AZ,United Kingdom}
\author{A. Letchford}
\author{C. Gabor}
\affiliation{STFC Rutherford Appleton Laboratory, Harwell Oxford, Didcot, OX11 0QX, United Kingdom.}
\date{\today}
\begin{abstract}
A non-invasive, compact laserwire system has been developed to measure the transverse emittance of an \hmm beam and has been demonstrated at the new LINAC4 injector for the LHC at CERN.
Light from a low power, pulsed laser source is conveyed via fibre to collide with the \hmm beam, a fraction of which is neutralized and then intercepted by a downstream diamond detector.
Scanning the focused laser across the \hmm beam and measuring the distribution of the photo-neutralized particles enables the transverse emittance to be reconstructed. 
The vertical phase-space distribution of a 3\,MeV beam during LINAC4 commissioning has been measured by the laserwire and verified with a conventional slit and grid method.
\end{abstract}
\maketitle
\section{\label{sec:intro}Introduction}
Modern proton driven accelerator applications, such as neutron spallation sources and high energy hadron colliders, demand increasingly higher beam currents.
A common solution is to inject \hmm ions via the charge exchange process, which overcomes the limit imposed by Liouville's theorem in circular machines~\cite{dimov1996}.
Such high beam currents present a challenge for conventional, invasive beam diagnostics, which may be damaged by the high beam powers.
Continuous online monitoring of the beam parameters also requires diagnostics that has minimal influence on the beam.
For this purpose, a laserwire provides an inherently indestructible and essentially non-invasive probe, that replaces the mechanical counterpart, such as a wire or slit.
The laser neutralizes a small fraction of the \hmm beam via the photo-detachment process, generating free electrons and \hz atoms.
Scanning the laser transversely across the particle beam and measuring either of the products of the interaction provides information on the beam properties.\\
Laserwire diagnostics for measuring an \hmm beam were originally built at Los Alamos National Laboratory~\cite{Cottingame1985, Connolly1992}, and subsequently developed at facilities including the Brookhaven National Laboratory LINAC~\cite{bnl93}, and the Spallation Neutron Source at Oak Ridge National Laboratory~\cite{Liu2010,Liu2012}, and more recently considered for future applications such as the Front End Test Stand (FETS) at Rutherford Appleton Laboratory, UK~\cite{Gabor2005, Gibson2013}.\\
Existing systems typically use lasers with peak powers above the megawatt level and free-space beam transport. Such laser systems need frequent maintenance, can be susceptible to stability and alignment issues, and also require a high level of safety measures. In contrast, the laserwire system presented in this paper has been designed to use a fibre laser with comparatively low pulse power in the kilowatt range. By transporting the laser light in an optical fibre to the interaction region the instrument is compact, has a stable output in terms of spatial and temporal properties and is unaffected by mechanical vibrations or misalignments. This has been demonstrated in \cite{Liu2013}, with a picosecond laser used to measure the longitudinal beam profile. The instrument developed for the LINAC4 at CERN will be used to measure the transverse beam emittance.\\
LINAC4 is currently being commissioned in the framework of the LHC (Large Hadron Collider) Injector Upgrade (LIU)~\cite{Hanke_LIU}. It will replace the existing proton accelerator, LINAC2, as the injector to the Proton Synchrotron Booster (PSB). After its completion expected by the end of 2016, the 90\,m long linac will produce an \hmm beam with mean current of 40~mA and energy of 160\,MeV \cite{TDR2006}. The transverse emittance of the LINAC4 beam must be monitored during operation to optimize injection into the PSB. Since conventional invasive diagnostics are unsuited to continuous online monitoring of the LINAC4 beam, instead, the development of a non-invasive laserwire instrument was proposed~\cite{Hofmann2013}. To achieve this goal, a collaboration between the CERN Beam Instrumentation group and the UK FETS beam diagnostics team was established~\cite{Gibson2013}. As the first step in this ongoing development, a laserwire emittance scanner has been prototyped and tested at the 3\,MeV beam energy during the first commissioning phase of LINAC4.\\
This paper describes the configuration of the laserwire prototype and presents first results from transverse emittance measurements of the 3\,MeV beam. The instrument design is first reviewed in Sec.~\ref{sec:design}, which includes simulation studies that instruct the system requirements. In Sec.~\ref{sec:exp_setup} the installed setup is detailed and the essential components are characterized in Sec.~\ref{sec:calibration}. The emittance measurement results are presented and compared with a conventional diagnostic in Sec~\ref{sec:results}. Finally, the paper provides an overview of future developments.
\section{\label{sec:design}Instrument design}
\subsection{\label{sec:concept}Principle of operation}
\begin{figure} [!hb]
\begin{center}
\includegraphics{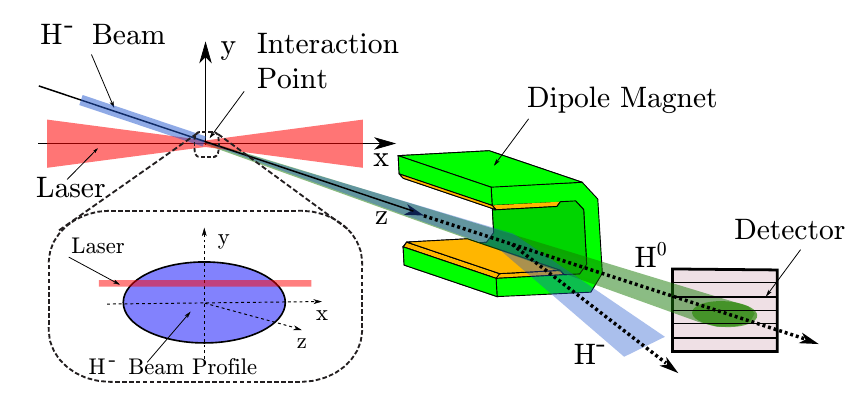}
\caption{Schematic principle of a laserwire emittance scanner. A fraction of the \hmm particle beam collides with photons of a focused laser beam. The neutralized \hz are detected by a downstream detector after being separated from the main beam by a dipole magnet. By measuring the \hz profile at the detector plane for different laser y-positions, the angular spread and ultimately the transverse phase-space can be reconstructed.}
\label{fig:concept}
\end{center}
\end{figure}

The operational principle of a laserwire instrument is to cross an accelerated beam of particles with a laser beam, such that the charged particles and photons can interact in a finely controlled overlap region, as shown schematically in Fig.~\ref{fig:concept}. The fundamental process exploited for \hmm beam diagnostics is photo-detachment; the weakly bound outer electron in the \hmm ion is permanently ejected if a sufficiently energetic photon is absorbed. The binding energy for the outermost electron is very low (0.75\,eV), and so is detached by photons with a wavelength of less than 1.65\,$\mu$m. Only a small transverse section of the particle bunch is sampled, as the laser is stepped through a series of y-positions. The distribution of the particles neutralized (\hz atoms) at each laser position is measured by a downstream detector, after being separated from the main beam by a bending dipole magnet. Merging each measured \hz profile with the vertical laser position provides the necessary information to reconstruct the transverse emittance of the \hmm beam. In our setup we conducted measurements only the vertical plane. The horizontal plane could be obtained likewise with the laser and detector setup 90$^{\circ}$ rotated.\\

\begin{figure}[!htb]
\begin{center}
\includegraphics{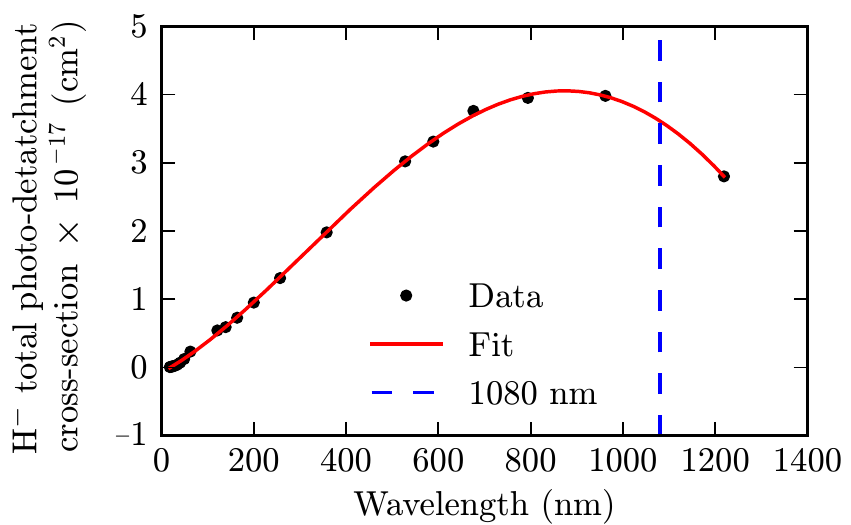}
\caption{Cross-section of the photo-detachment process of a non-relativistic \hmm ion. The dashed line indicates the selected laser wavelength for the emittance measurements at 3\,MeV.}
\label{fig:cross}
\end{center}
\end{figure}

The dependence of the photo-detachment cross-section on the wavelength of the incident photon~\citep{detach} is shown in Fig.~\ref{fig:cross}. The probability of ejecting one electron by interacting with a laser photon in the visible or near infrared region is generally high so the signal produced by the laserwire will be robust even when relatively low power laser pulses are used. However, because of the very low binding energy of the outer electron in the \hmm, the particles can be easily neutralized also by means of other processes such as black body radiation, magnetic fields or collisions with residual gas atoms, the latter being the dominant source of background \cite{cheymol2011}. In the next subsection, the processes of photo-detachment and collisions with the residual gases are considered quantitatively to predict the expected ratio between signal and background.
\subsection{Theoretical models and simulations}
A theoretical framework of the laserwire interaction was developed to evaluate the expected performance of the laserwire instrument and establish the range of key parameters required for the laser source and detection system. The theoretical framework consists of a model of the laser photo-detachment signal, a model of the residual gas background and a comparative simulation.
\subsubsection{Photo-detachment model}
The probability that the outer electron is stripped from an \hmm ion during exposure to laser radiation is modeled by~\cite{sns2002}:
\begin{equation}\label{eq:stripping_probability}
\mathbb{P}_{Laser}=1-\exp\left(-\sigma(E_{CM})\rho(x,y,z)t\right),
\end{equation}
where $\sigma(E_{CM})$ is the photo-detachment cross-section, $\rho(x,y,z)$ is the laser photon density and $t$ is the time of interaction. The cross-section in~Eq.~\ref{eq:stripping_probability} depends on the center-of-mass energy, \begin{equation}\label{eq:cm_energy}
E_{CM}=\gamma E_L\left(1-\beta cos\Theta_L\right),
\end{equation}
which is a function of the laser energy $E_L$, the Lorentz factor  $\gamma=1/\sqrt{1-\beta^2}$, the relativistic velocity ratio of the \hmm ions $\beta=v/c$, and the angle $\Theta_L$ between the laser and particle beams. In the case of the 3\,MeV \hmm beam at LINAC4, the Lorentz factor is small enough that relativistic corrections have negligible impact on the cross-section. For the final laserwire system that will measure a 160\,MeV beam, the effective value of the selected laser wavelength will be relativistically shifted close to the peak in the photo-detachment cross-section of Fig.~\ref{fig:cross}.

The model simulated the stripping interactions that occur when an input distribution of \hmm ions cross the laser beam. Equation~\ref{eq:stripping_probability} was applied to each \hmm ion, taking into account the time spent by the ion as it traverses the varying photon density of the laser beam. Thus the portion of ions stripped by the laser was determined, as detailed in~\cite{Hofmann2013}. The simulations were performed for one laser pulse travelling in the $x$-direction that is geometrically centered in $y$ on the \hmm beam. The trajectories of \hz particles produced in this overlap region were propagated 3\,m downstream to the detector plane and integrated, such that the expected measurable profile of particles could be plotted.
\subsubsection{Neutralization due to collisions with residual gas}
The probability of ejecting one electron from an \hmm ion by collision with a stray particle present in the vacuum pipe is
\begin{equation} 
\mathbb{P}_{Gas}=1-e^{-l/\lambda}\ {\rm with}\ \lambda=\frac{kT}{\sigma P};\,\sigma\sim\frac{1}{\beta^2_{rel}}.
\label{eq:gas_strip}
\end{equation}
This stripping probability $\mathbb{P}_{Gas}$ is calculated from the interaction length $l$, the beam pipe temperature $T$, and the pressure $P$ of the residual gas, which consists mainly of hydrogen. The dependence of the collisional detachment cross-section, $\sigma$, on the beam energy was derived from the literature~\citep{gas_lebt,gas_mev}.
\subsubsection{Signal and background simulations}
\begin{figure}[!b]
\begin{center}
\includegraphics{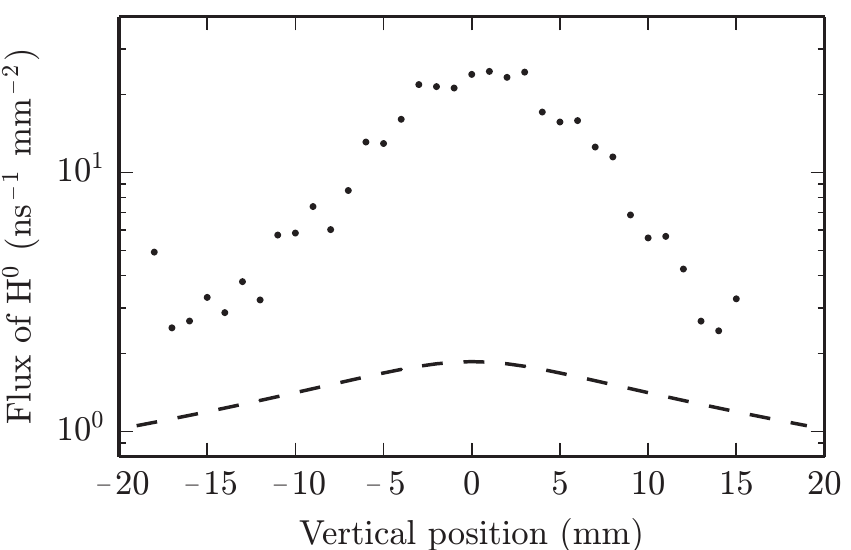}
\caption{Flux of \hz  arriving at the detection plane from laser stripping in the center of the \hmm beam - in average while laser pulse duration - (dots); \hz flux from the residual gas background - permanently while macro-pulse - (dashed line).}
\label{fig:laser_backg}
\end{center}
\end{figure}

The parameters used to calculate the distribution of photo-detached ions at the detector plane are listed in Tab.~\ref{tab:sim_laser}.
Of particular interest was the calculation of the signal produced by a laser with a relatively low peak power (in the kW range) that could be transported in optical fibre. The laser pulse duration was also chosen in a way that the produced signal could be realistically detected and digitized without bandwidth restrictions. The value for the \hmm beam current and the beam size were set to reflect real conditions of the LINAC4 machine during commissioning.\\
The simulations for the background distribution in the plane of the \hz detector were performed using gas pressures and beam dynamics data from the \hmm source to the bending magnet, through the accelerator components described in Sec.~\ref{sec:linac4},
Tab.~\ref{tab:conditions_backg} summarizes the parameters for the background simulation. $\mathbb{P}_{Gas}$ was calculated in 2\,cm intervals along the beam axis and the stripped portion was propagated to the \hz detector. 
\begin{table}[h]
\begin{center}
\caption{Photo-detachment signal simulation parameters.}
\label{tab:sim_laser}
\begin{tabular}{l c r}
\hline\hline
Parameter & Value & Units\\
\hline
Laser pulse energy          &67          &$\mu$J \\
Laser pulse length (2$\sigma$)          &106         &ns \\
Laser waist radius          &76          &$\mu$m \\
Ion beam current            &8.5        &mA \\ 
Vertical ion beam size      &1.01       &mm\\
Distance from the IP to the detector		&3.4    &m\\ 
\hline\hline
\end{tabular}
\end{center}
\end{table}
\begin{table}[htbp]
  \centering
  \caption{Residual gas background simulation parameters.}
    \begin{tabular}{lcccr}
    \hline
    \hline
    Parameter               &LEBT   &RFQ    &Chopper    &Units\\
    \hline
	Length                  & 2.0   & 3.0   & 3.7   & m  \\
    Residual gas pressure   & 20    & 10    & 1     & 10$^{-10}$\,bar  \\
    Min. aperture radius    & 50    & 3.54  & 6.2   & mm  \\
    Mean pulse current      & 13    & 11    & 8.5   & mA  \\
    \hline
    \hline
    \end{tabular}%
  \label{tab:conditions_backg}%
\end{table}
The results of both the simulations of signal and background are summarized in Fig.~\ref{fig:laser_backg}. 
The dotted curve in the plot shows the resulting \hz density in the detector plane produced by a laser pulse crossing the  3\,MeV \hmm beam center - in average while laser pulse duration - and the dashed curve corresponds to the flux of \hz produced by residual gas interaction. Due to the time structure of both signals being very different, they can be separated in frequency domain without difficulty. In fact, being the signal produced only during the interaction with the pulsed laser, it will have a similar duration, whereas the background, as it is generated all along the beam-pipe, will be composed of \hz s with kinetic energies ranging from 45\,keV to 3\,MeV and it arrives at the detector plane unbunched and with a duration comparable with the particle macro-pulse. 

\subsection{\label{sec:requirements}System requirements}
Analyzing the simulation result, summarized in Fig~\ref{fig:laser_backg}, one can derive requirements for the laser and detector system.\\ Tab.~\ref{tab:laser_req} gives an overview of the requirements for the laser system. The wavelength is for our system not very critical, since the cross-section is very broad (see Fig.~\ref{fig:cross}). Therefore we aimed for 1064\,nm laser where a wide choice of lasers are commercial available. To reach a signal level which is significant above the background level one needs a laser with peak power in the kilowatt range. To relax the bandwidth requirements for the detector, a laser pulse-length of at least 10\,ns is required. With an repetition rate of more than 10\,kHz, synchronized to the LINAC4 timing one can sample the 400\,$\mu$s \hmm pulse multiple times. \\
To realize a laserwire scanner, a nearly constant diameter of the laser beam which is a factor of 10 smaller than the \hmm beam diameter is the pre-condition. Taking the \hmm beam $\sigma$ of 1...3\,mm, a laser beam with a beam quality factor ($M^2$) smaller than 3 is needed. E.g. in our simulation we used an laser with $M^2$ = 1.8 and achieved a waist radius of 76\,$\mu$m with a Rayleigh length of 6.5\,mm. 
This will be discussed more in detail in section~\ref{sec:inst_setup}\\

\begin{table}[h]
\begin{center}
\caption{Requirements for the laser system}
\label{tab:laser_req}
\begin{tabular}{lcr}
\hline\hline
Parameter                           &Requirement        &Units\\
\hline
Wavelength                          & 900 $\pm$ 300     &nm\\
Pulse peak power                    & 1...10            &kW\\
Pulse length                        & 10...100          &ns\\
Repetition rate                     & $>$ 10            &kHz\\
Synchronisation to LINAC4           & Required          &--\\	
Beam quality factor ($M^2$)         & $<$ 3             &--\\
\hline\hline
\end{tabular}
\end{center}
\end{table}
Regarding the \hz detector the time resolution is a key-factor. It needs to resolve a laser pulse so that the signal is not dominated by residual gas background but laser stripped \hz. A bandwidth of min. $20$\,MHz is required to record the beam pulse with negligible distortion \cite{hofmann15}. To achieve a reasonable electrical signal the detector also needs to have internal gain. This requirement follows from simulation results shown in Fig.~\ref{fig:laser_backg}. As the \hz density at the detector will be in the order of 10~$\hz ns^{-1}\,mm^{-2}$, it corresponds to only nanoampere currents.\\
On the same time, the background flux of approx. 1~$\hz ns^{-1}\,mm^{-2}$ can create significant damage in detector materials integrated over the LINAC4 pulse length of 400\,$\mu$s. For this reason the radiation resistance of the detector is also an important factor.
\section{\label{sec:exp_setup}Experimental setup}
\subsection{\label{sec:linac4}Overview of LINAC4}

\begin{figure}[!hb]
\begin{center}
\includegraphics{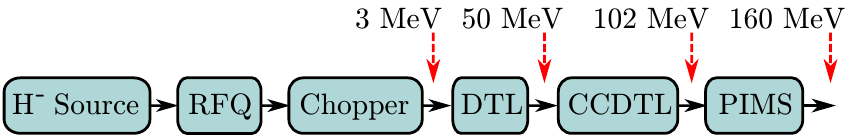}
\caption{LINAC4 block diagram.}
\label{fig:l4}
\end{center}
\end{figure}

The final layout of the LINAC4  will consist of a 45\,keV \hmm ion source, a radio frequency quadrupole (RFQ), which accelerates the beam to 3\,MeV, and a chopper. A drift tube linac (DTL), a cell-coupled drift tube linac (CCDTL) and a Pi-mode structure (PIMS) will be then used to accelerate the beam to its final energy of 160\,MeV\,\cite{l4}. During the LINAC4 construction, the beam commissioning is taking place in stages and at every stage the beam has to be fully characterized. A schematic block diagram of the machine is shown in Fig.~\ref{fig:l4} and the design parameters of the LINAC4 are summarized in Table\,\ref{tab:l4params}.
\begin{table}[htb!]
\caption{\label{tab:l4params} LINAC4 Parameters}
\begin{tabular}{l c c r}
\hline \hline
Parameter             & Symbol             & Value               & Units\\
\hline
Overall linac length        			& $\text{L}$            			& 90    & m \\
Output energy               		& $\text{E}$            			& 160   & MeV \\       
Bunch frequency       		& f$_{\text{bunch}}$    	& 352.2 & MHz \\
Pulse length           		& t$_{\text{pulse}}$    		& 400   & $\mu$s\\
Pulse repetition rate          & t$_{\text{rep}}$       & 1.2   & s\\
Average pulse current       		& I$_{\text{pulse}}$    		& 40    & mA\\
Beam transverse emittance		& $\varepsilon$         		& 0.4   & $\pi$~mm~mrad\\
\hline\hline
\end{tabular}
\end{table}
\subsection{\label{sec:3mev_setup}Diagnostic setup at 3\,MeV}
\begin{figure*}[!tbh]
\centering
\includegraphics*[width=2\columnwidth]{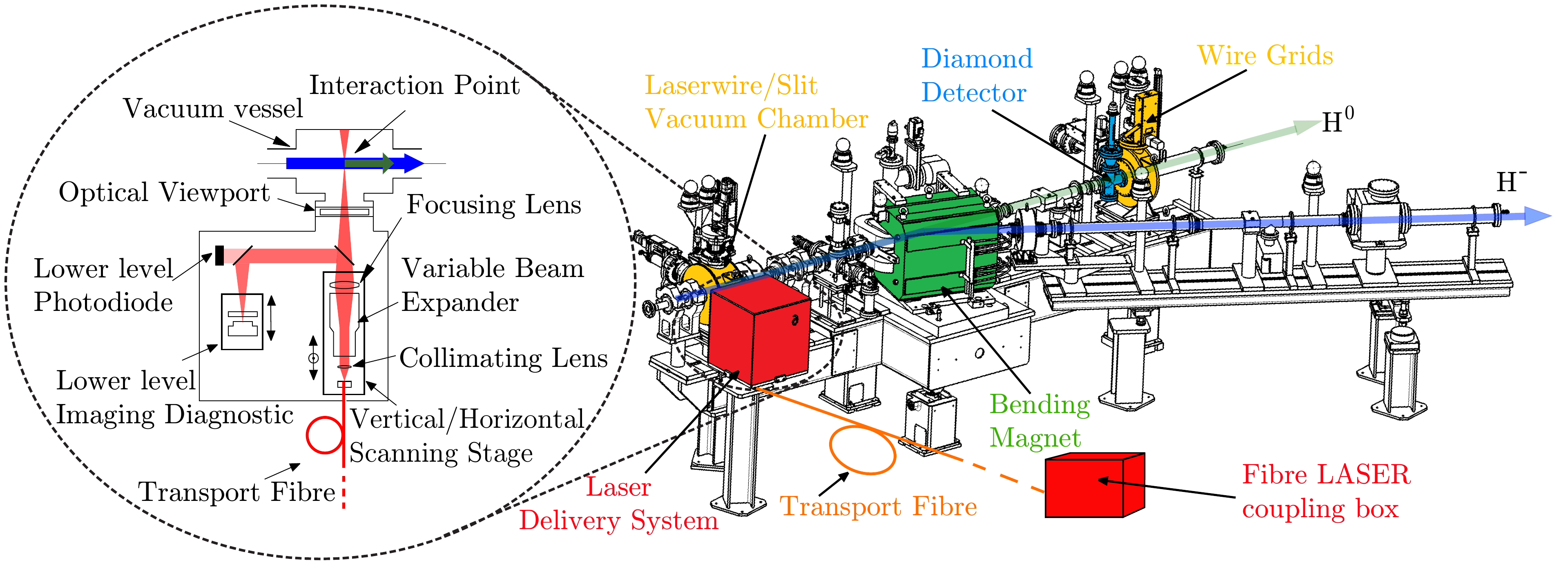}
\caption{Layout of the \SI{3}{MeV} diagnostics test bench as it was installed after the Chopper, including the operational emittance meter (slit and grids) and the laserwire emittance scanner prototype.}
\label{fig:test_bench}
\end{figure*}

The laserwire system was integrated into a movable test bench that was used to measure the beam parameters after the chopper, during the commissioning phase at 3\,MeV. The test bench also contained various diagnostic tools such as beam position monitors, beam current transformers (BCT), slit and grid emittance scanner etc. for full characterisation of the beam at different commissioning stages. The schematic diagram of the test bench is illustrated in Fig.\,\ref{fig:test_bench} (right hand side).\\
The laser and \hz monitoring subsystems, which will be described in detail in the next sections, have been installed as close as possible to the slit and wire grids respectively in order to facilitate the comparison between the two methods. The spectrometer dipole in between the slit (and laser) and the wire grids (and \hz detector) was designed to steer the beam (during periods not used for slit-grid measurements) into the spectrometer line for absolute and relative energy measurements via horizontal profile measurements at the end of this line. Powering such a dipole allows separating the stripped \hz (to be measured in the straight line by the diamond detector) from the un-stripped \hmm particles and thus performing the laserwire measurements.\\
Simulations of the beam envelope with nominal quadrupole settings resulted in an expected beam size in the plane of the laser of $\sigma_x = 2.6$\,mm and $\sigma_y = 1.4$\,mm.
\subsection{\label{sec:laser_setup}Laser system and optical layout}
The laser is a Q-switched, diode pumped, Yb doped fibre Master Oscillator Power Amplifier (MOPA) manufactured by ``Manlight S.A.S'' (Lannion, France), model: ML-30-PL-R-TKS. The oscillator generates pulses with approximately 80~ns width (FWHM) at wavelength of 1080~nm and a repetition rate selectable between 30 and 100~kHz.

The fibre amplifier is pumped by a laser diode that can run either in continuous wave (CW) or can be driven by an external signal with a repetition rate of up to 5~kHz and an adjustable duty cycle, which enables the train of amplified pulses to be synchronized to an external source. The maximum average power of the laser system is approximately 28~W when pumped in continuous wave (CW) regime and the maximum pulse peak power about 8.5~kW.

\begin{figure}[!htb]
\begin{center}
\includegraphics{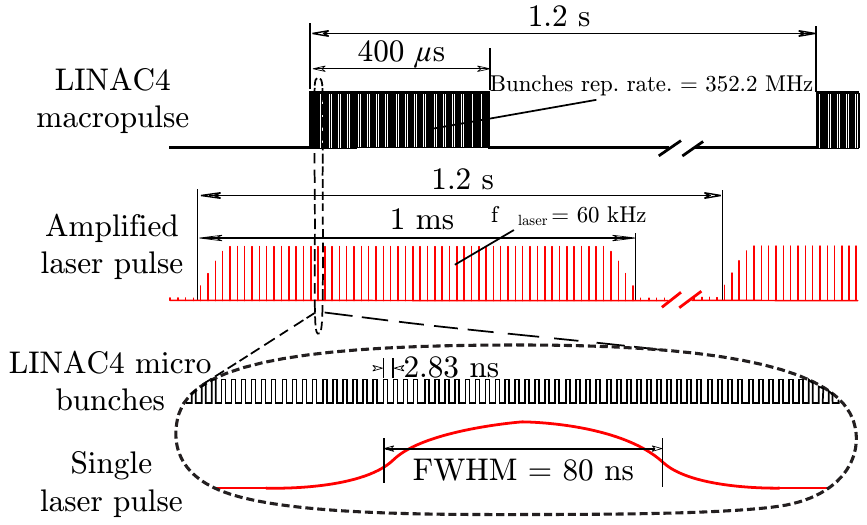}
\caption{Timing diagram of the laserwire system.}
\label{fig:timing}
\end{center}
\end{figure}

In order to optimize the duty cycle of the laser, and therefore use a lower average power, the pump diode was synchronized to the LINAC4 macropulse. Also, the pulse duration was set to 1 ms in order to comfortably overlap the laser amplified pulses with the 400 $\mu$s long macropulses. A detailed overview of synchronization between the laser pulses and the accelerator bunches is shown in Fig.~\ref{fig:timing}, including a zoomed view of the overlap between the single laser pulse and the ions micro-bunches.

The transport system designed to convey the laser beam to the interaction point is based on a large mode area (LMA) optical fibre. The schematic of the optical system is shown in Fig.\,\ref{fig:test_bench} (left hand side). The coupling and focusing optics were mounted on breadboards and enclosed in two separate interlocked boxes. In the coupling box, the output from the Q-switched MOPA fibre-laser is coupled into a 5~m long optical LMA fibre with a core diameter of 20~$\mu$m. In order to continuously monitor the laser power participating to the photo-neutralization a fast photodiode with sub-nanosecond risetime is set inside the coupling box to pick up a parasitic signal. The laser delivery system is setup just in front of the beam-pipe, which is accessible to the laser via  a anti-reflexion coated vacuum window. The output face of the fibre together with the focus optics are set on a small plate and mounted on a stack of two automated translation stages with 50 mm range and $\sim$1~$\mu$m step resolution, one set to scan the laser beam vertically (scanning stage or Y-stage) and a second one to allow longitudinal positioning of the laser focus within the vacuum pipe (longitudinal stage or X-stage). The focus optics consist of collimation lens with 6.24~mm focal length, a remotely controllable variable beam expander that has a range of magnification adjustable between 1x and 8x and a focusing lens with focal length of 500 mm that focuses the laser radius to $\sim$75~$\mu$m at the interaction region, which is much smaller than the particle beam size to be scanned.

A secondary optical path is also installed within the focusing box. It is setup in a way that when the Y-stage is set at its lower position, the laser beam is directed towards an optical window that reflects a small portion of light into a CCD camera set on a translation stage so that the laser spatial characteristics can be measured on demand. The larger part of light that is transmitted through this optical window is detected by a photodiode for peak power and pulse duration measurements.

\subsection{\label{sec:diamond_setup} Diamond detector system}
\begin{figure}[!ht]
\begin{center}
\includegraphics[width = 7cm]{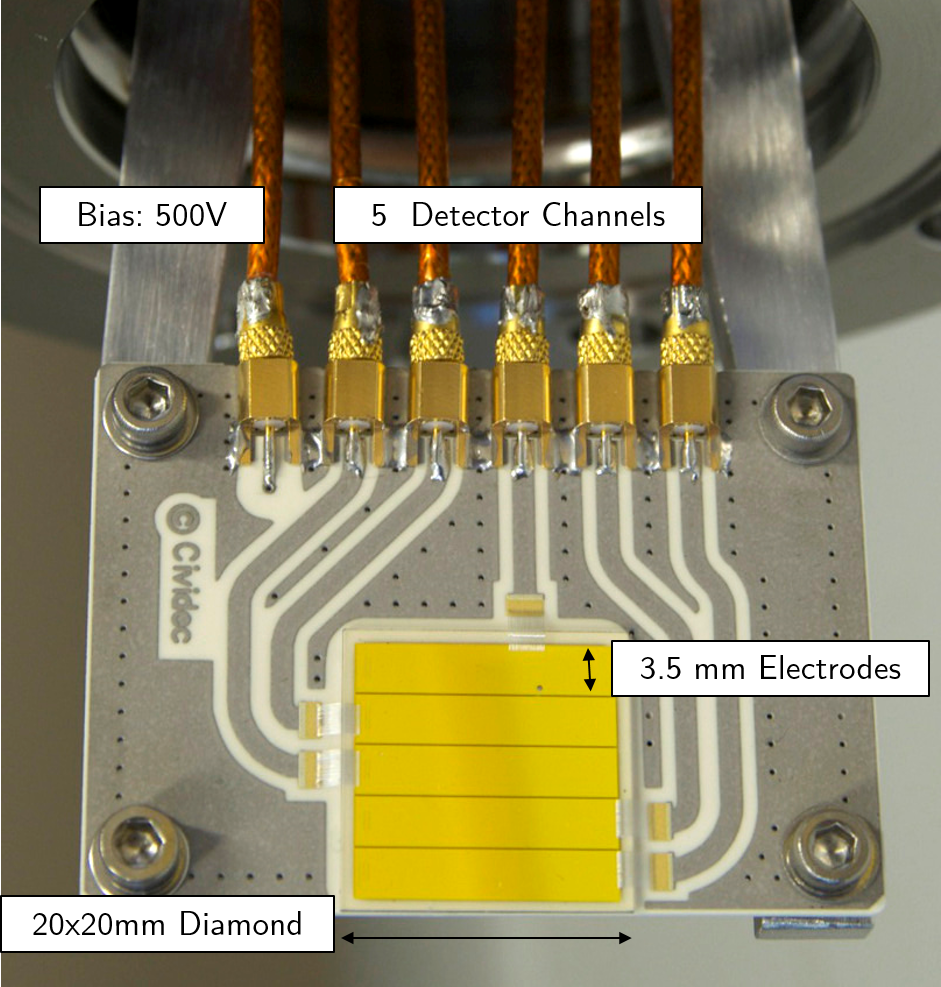}
\caption{pCVD diamond detector with 5 strip electrodes \cite{cividec}.}
\label{fig:diamond}
\end{center}
\end{figure}

In section~\ref{sec:requirements} the requirements for a suitable \hz detector were determined. After analysis of different types of detectors we choose a  polycrystalline (pCVD) diamond strip detector for the measurement campaign. This kind of detector is able to measure even single particle events due to its internal gain of about $10^4$ electrons per penetrating \hz. The response time in the sub-nanosecond range permits to resolve an arriving laserpulse (FWHM~=~80~ns) with sufficient time resolution. Finally, the radiation hardness of approx. $10^{15}\,proton/cm^2$ comparing to $10^{14}\,proton/cm^2$ for silicon \cite{si_vs_dia} is essential.\\
A photo of the detector is shown in Fig.~\ref{fig:diamond}. The 20~x~20\,mm, 500\,$\mu$m thick pCVD diamond disc is mounted on a ceramics printed circuit board (PCB) while the five 0.2\,$\mu$m thick aluminium electrodes on the front side of the detector are bonded to the circuit paths. On the PCBs backside a bias voltage of 500\,V is applied to readout the charge created in the diamond bulk. Capacitors (1\,$\mu$F) in parallel to the detector rapidly recharge the diamond when electrons are read out.\\
The PCB with the detector is fixed to a fork-shaped support-frame situated inside a 4-way crosschamber. The setup is mounted on a stepping motor which provides vertical movement in a range of 80\,mm  with 150\,$\mu$m resolution.\\
The signals need to be pre-amplified directly after the vacuum feed-through due to an unexpected low sensitivity of the detector. The effect causing this will be discussed quantitatively in chapter~\ref{dia_sig}. We were using AC-coupled linear amplifiers with 46\,dB gain and 100\,MHz bandwidth. Thereafter the signal from the diamond channels were digitized with a 1 GS/s oscilloscope.
\section{\label{sec:calibration}System characterisation and measurements}
\subsection{\label{sec:inst_setup}Laser characterisation}
\begin{figure}[!htb]
\begin{center}
\includegraphics[width=1\columnwidth]{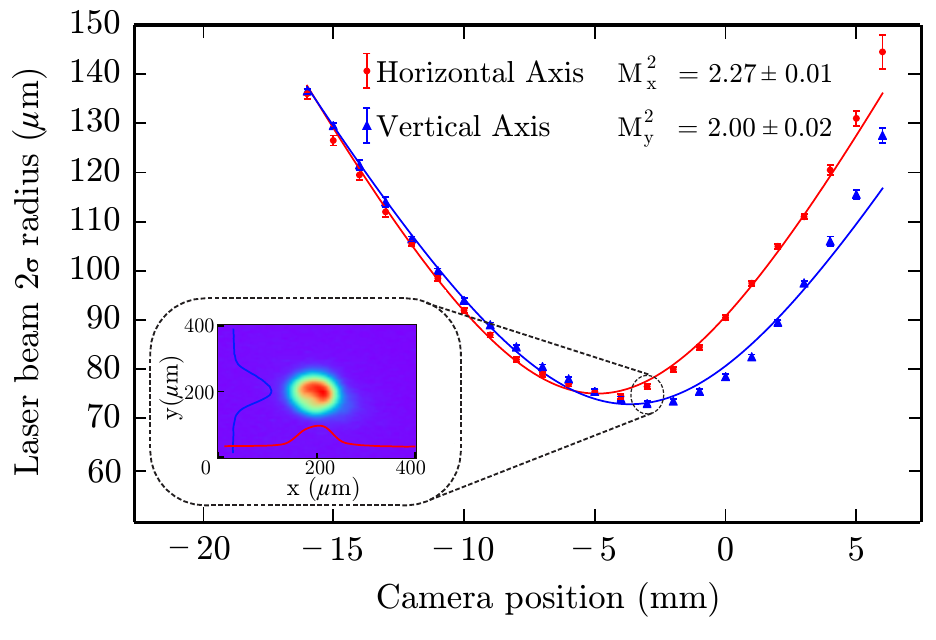}
\caption{Measured 2$\sigma$ width of the laser beam focused by a lens with a focal length $f$ = 500~mm. At the bottom left is shown the laser spot image taken at a position of -3mm, where the minimum vertical size is obtained.}
\label{fig:m2}
\end{center}
\end{figure}

As the laser beam is used as a probe to scan the particle beam, a full characterisation of its properties is required.
To measure the spatial characteristics of the laser beam we used a CCD camera which was set on a motorized translation stage in our lower level diagnostic optical path (see laser delivery system in Fig.~\ref{fig:test_bench}). We moved the camera in steps of 1~mm and recorded the beam spot size variation along the longitudinal axis. Fig.~\ref{fig:m2} shows the laser spot size at the focal plane of the focusing lens. The horizontal and vertical laser diameters extracted from the images are plotted along the direction of propagation in Fig.~\ref{fig:m2}. The two sets of data were fitted with the laser propagation formula for quasi-Gaussian beams,
\begin{equation}
\label{eq:m2} 
W(z)=W_{0}~\sqrt{1+(z \frac{\lambda~M^{2}}{\pi~W_{0}^{2}})^{2}}
\end{equation}
where W(z) is the laser spot radius (distance from the center of the distribution to the position where the intensity drops by a factor $e^{2}$), $W_{0}$ is the minimum laser waist, $\lambda$ is the laser wavelength and $M^{2}$ is a factor which describes the quality of the real beam compared to an ideal $TEM_{00}$ Gaussian beam (for which $M^{2}=1$) \cite{laser_phy}.
For our application the results in the vertical plane are by far more important. They define the resolution for the sampling (in analogy to the slit size). The resulting value for $M^{2}_y$ is 2.0, the laser waist $W_{0y}$ at the focal plane of the lens is approximately 75$\mu$m (corresponding to a $\sigma$ of 37.5$\mu$m) and the Rayleigh length is about 8\,mm. Summarizing the characteristics and comparing to the \hmm beam the laser beam can be approximated as a thin cylinder with constant diameter ($\diameter = 150 \mu$m).\\
Regarding the characterisation of the longitudinal parameters of the laser pulse we measured the pulse shape before and after the fibre propagation with fast photodiodes. Furthermore we measured the average power before and after the fibre to check the coupling and transport efficiency of the fibre transport. The efficiency of the fibre delivery was found to be about 70\,\%\,$\pm\,5\%$ with higher efficiency at lower laser power.\\
For the emittance measurements the laser beam pulse energy interacting with the \hmm beam was measured to be 154\,$\pm\,6\,\mu$J.\\
\subsection{\label{sec:diamond_signal}Diamond detector signal examination}
\label{dia_sig}
\begin{figure}[!t]
\begin{center}
\includegraphics{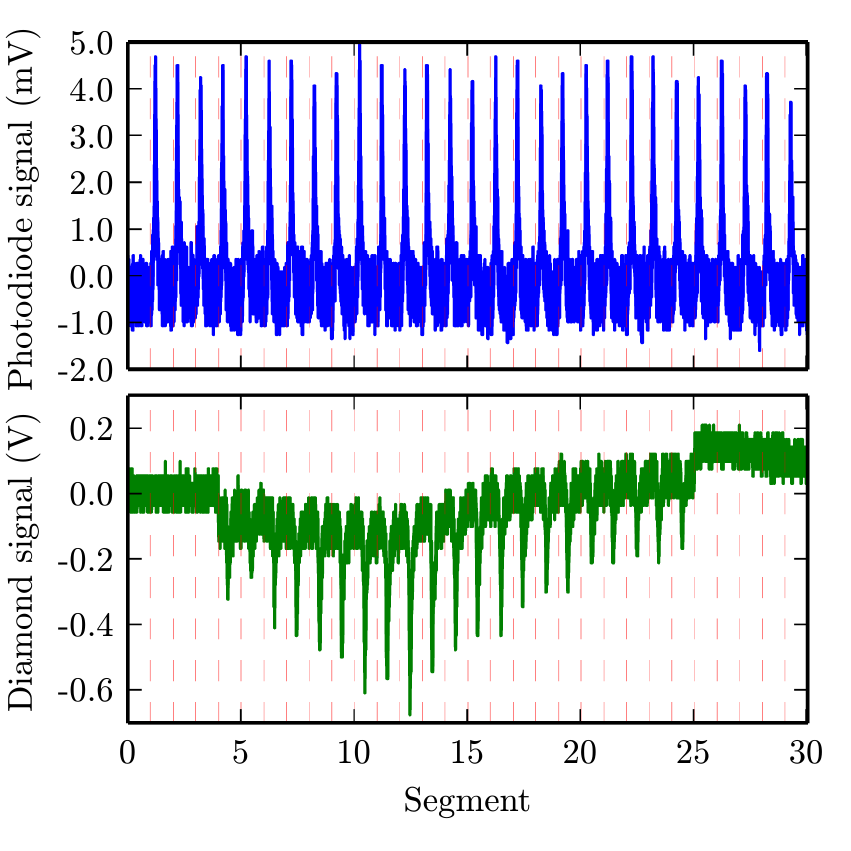}
\caption{Comparison of laser pulses recorded with a photodiode and the amplified signal at the diamond detector.}
\label{fig:detector_pd_signal_trace}
\end{center}
\end{figure}

To test the signal response of the diamond detector, laser and detector were positioned in order to maximize the signal in one specific channel of the diamond detector.\\
Fig.~\ref{fig:detector_pd_signal_trace} shows in the upper plot the amplified laser pulses, interacting with the \hmm beam, recorded with a photodiode in the coupling box. In the lower plot, the resulting signal of an amplified channel of the diamond detector during a \hmm beam pulse is displayed. This signal corresponds to one laser position and one diamond position and each segment represents a 1~$\mu$s time-interval.\\
Comparing the continuous laser pulses with the signal on the diamond detector, one finds a clear correlation just between the segments 4 and 24 where the 400~$\mu$s long \hmm macropulse can interact with the laserpulses. In this timeframe the \hz created by the laser interaction form sharp peaks in the diamond signal. Furthermore the \hz background originating by collisions with the residual gas molecules cause an offset on the diamond.

\begin{figure}[!b]
\begin{center}
\includegraphics{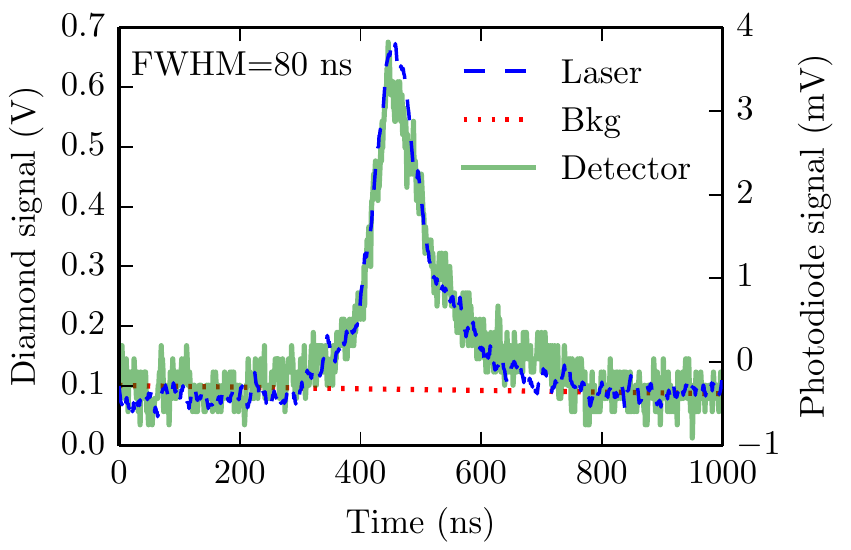}
\caption{Zoom on one of the diamond signals of Fig.~\ref{fig:detector_pd_signal_trace} (solid - inverted), compared to the corresponding laser pulse (dashed). Dotted trace -- linear fit of the background.}
\label{fig:pulses}
\end{center}
\end{figure}

\begin{figure}[!htb]
\begin{center}
\includegraphics{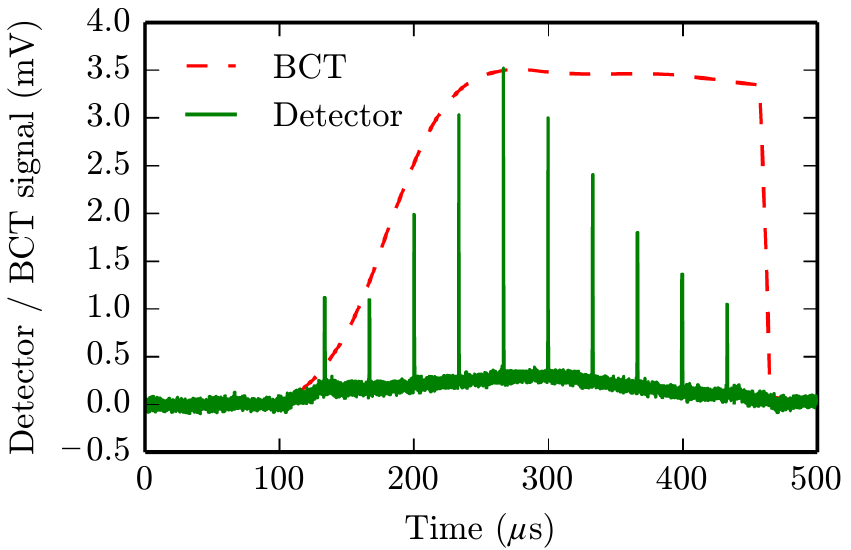}
\caption{Unamplified signal from a diamond detector channel sampling the \hz particles for a given laser position ($E_{pulse}~=~67~\mu J; f_{pulse} = 30~kHz$), as function of time along the 3\,MeV \hmm beam pulse. This is compared to the beam current measured with a beam current transformer (BCT) downstream the laser station.}
\label{fig:diamond-vs-bct}
\end{center}
\end{figure}

Fig.~\ref{fig:pulses} shows a zoom to one of the pulses where the diamond signal (solid) and the laser pulse (dashed) are overlaid. The distinct agreement in the shape of the signals demonstrates the linear relation between laser-power and output signal of the diamond detector.\\
To characterize further the detector response, a comparison was done between the \hmm beam current recorded shortly behind the laser interaction and the unamplified diamond detector raw signal. Fig.~\ref{fig:diamond-vs-bct} shows the relation of both signals. The slight increase of the floor of the diamond signal can be explained by the \hz atoms created by residual gas collisions. The ratio between the peak signal and baseline value is fairly constant (between 10 and 11). Given the uncertainties of the simulation input (e.g. gas pressure, detector time response etc...), this value compares well to the simulations (see Fig.~\ref{fig:laser_backg}).

In addition the time response of the diamond signal was compared with the BCT signal all along the LINAC4 macropulse. Even though in some cases the signal from the diamond detector reproduced well the beam current evolution~\cite{linac14}, in most cases, as presented here, the agreement was poor in the second part of the beam pulse.\\
Even though such an occurrence, obviously affecting the overall measurement accuracy, has not been fully understood yet, it can be well related to the implantation of \SI{3}{\mega\eV} protons into the diamond bulk, after approx. 50\,$\mu$m of detector material. Simulation results give more than $10^{7}$ protons implanted by the background for one LINAC-pulse. The variation of the electrostatic field inside the diamond could thus lead to a decreased Charge Collection Efficiency (CCE) during the LINAC macropulse. This might also explain the signal amplitude of just a few mV, which is two orders of magnitude lower than expected by calculations not considering the effects of the proton implantation \cite{hofmann15}. This disturbance is supposed to be negligible for higher beam energies (starting from \SI{12}{\mega\eV}) for which the protons have high probability to traverse and escape the diamond bulk without being stopped. First measurement results at the 12\,MeV beam show first evidence confirming this hypothesis.\\

\section{\label{sec:results}Profile and emittance measurement results}
\subsection{\label{sec:data}Analysis method}
To extract the transverse emittance value from the raw data recorded with the diamond detector an analysis routine was developed and applied to the gathered data. The first action is to define the segments with useful data in the digitized signal (e.g. Seg. 6-23 in Fig.~\ref{fig:detector_pd_signal_trace}). Hereafter each segment is treated individually. At first a finite impulse response (FIR) low-pass filter (3\,dB cutoff at 22\,MHz) is used to suppress high-frequency noise. Now a time-interval is defined, which separates the laserpulse from the background (in Fig.~\ref{fig:pulses} e.g. 300...700 ns). By fitting linearly the data outside of this window a subtraction of the diamond signal from the background can be conducted. Finally the data in the time-interval of the laserpulse can be integrated.\\ 
To get a mean emittance value along the duration of the LINAC4 macropulse, the integrated values for the selected segments are averaged. The resulting value corresponds to one sample in the transverse phasespace. The referring $y/y'$ values are calculated by the current detector and laser position.\\
The RMS transverse emittance can finally be calculated by using these samples in the phasespace as shown in Eq.~\ref{eq:em_formula}.

\begin{equation}\label{eq:em_formula}
\epsilon_{y} = \sqrt{\langle y^2 \rangle \langle y'^2 \rangle -\langle yy' \rangle^2}
\end{equation}
To extract the vertical beam profile from this data, the values at different $y'$ positions can be integrated for each $y$ coordinate. 

\subsection{\label{sec:emittance}Profile and emittance reconstruction and comparison with slit~\&~grid system}
\begin{figure}[!htb]
\begin{center}
\includegraphics{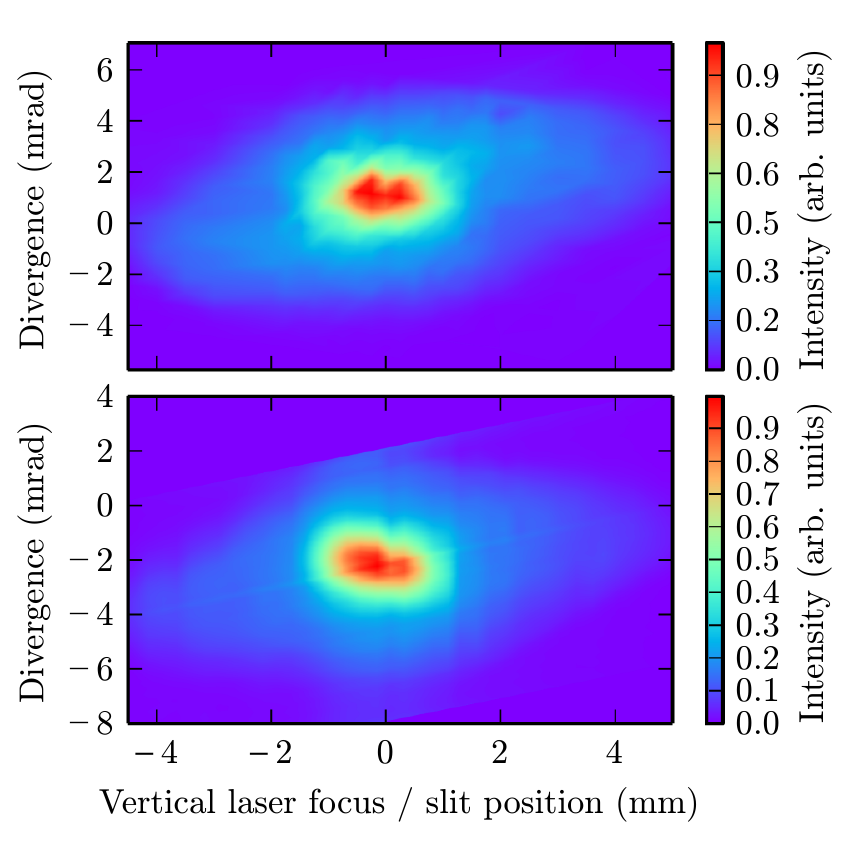}
\caption{Vertical transverse phase-space of the 3 MeV \hmm beam measured by laserwire (top) and slit and grid method (bottom).}
\label{fig:phase_space}
\end{center}
\end{figure}

In order to measure vertical transverse emittance the following technique has been used. The laser focus was moved vertically across the ion beam with variable step size (0.5~mm at the tails and 0.25~mm around the peak) in order to accurately sample the shape of the distribution. At each laser position, the diamond detector was scanned across the beamlet of neutral particles with steps of 1.81~mm in order to increase the angular resolution. Each detector scan corresponding to certain laser focus position was stored in separate data file. The data from these files then were analysed using the same analysis method as described above. The integrated and averaged signal then was used to plot the phase-space distribution.\\
The laserwire results were verified with the independent measurements from conventional slit and grid emittance meter. The slit, installed at the same point as the laser IP was scanned across the ion beam in vertical direction. At each vertical position, well defined with respect to the ion beam center, the slit selects a narrow slice of the ion beam. The distribution of the particles transmitted through the slit was measured after 3.4~m drift space by two wire grids~\cite{slitgrid}. Same as in case of laserwire, by scanning the slit vertically across the ion beam and measuring the profiles of particle beamlets passed through the slit the whole phase-space was reconstructed. Measured phase-space distributions for 3~MeV setup obtained with laserwire and slit and grid methods are presented in Fig.~\ref{fig:phase_space}.

\begin{table}[htbp]
  \centering
  \caption{Comparison of normalized emittance and Twiss parameters obtained with both methods}
    \begin{tabular}{lcc}
    \hline
    \hline
                                   &Slit-Grid   &Laser-Diamond\\
    \hline
	Norm. emittance [$\pi$ mm mrad]   & 0.242   & 0.215\\
	Beta [m]                        & 0.906   & 0.848\\
	Gamma [$m^{-1}$]                & 1.16    & 1.328\\
	Alpha                           & -0.225  & -0.348\\
    \hline
    \hline
    \end{tabular}%
  \label{tab:twiss}%
\end{table}
\begin{figure}[!htb]
\begin{center}
\includegraphics{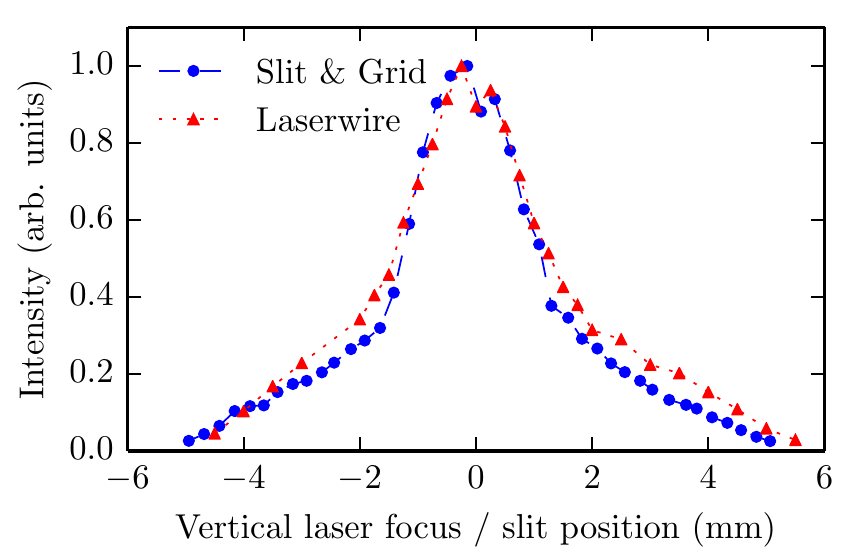}
\caption{Vertical profile of the \hmm beam measured by laserwire (dotted curve) and slit and grid (dashed curve) methods.}
\label{fig:profiles}
\end{center}
\end{figure}
As one can see from the plot, despite the lower spatial resolution of the laserwire data in this first test, the ellipse size and orientation in both pictures is in a good agreement. Vertical position of centroids on both pictures are slightly different. It can be explained by the fact that the exact position of the laser focus and the detector center was not perfectly aligned with respect to the center of the ion beam.\\
Table~\ref{tab:twiss} shows a comparison of the obtained values for the normalized RMS emittance and the twiss parameters. Both methods show reasonable agreement. The differences can be explained by several effects. The implantation effects of the protons in the diamond bulk might have caused some non-linearities in the detector response. Furthermore for the quadrupol settings used, a small percentage of particles might have missed the diamond detector, as it is smaller than the wire-grid. Finally, the noise suppression excluded all values in phase space which are below a certain threshold corresponding to a percentage of the maximum amplitude. Some dependency on the applied threshold also has an influence on the results shown in Table~\ref{tab:twiss}.\\
In order to compare as well the vertical beam profiles, projections of the phase-space distributions were calculated. The vertical profile of the ion beam for both laserwire and slit and grid measurements is presented in Fig.~\ref{fig:profiles}.
As one can see, the beam core is in a very good agreement. Values at the tails are slightly bigger when measured with the laserwire system. Also, it should be noted that the beam profiles obtained with two different methods have a non-Gaussian shape. The fact that the vertical profile of the ion beam has a non-Gaussian shape regardless of the measurement methods indicates that such shape of the beam is caused by the processes of beam generation at the source rather than a measurement error.

\section{\label{sec:conclusions}Summary and outlook}

In the last 2 years the development of a laserwire emittance scanner was accomplished. In a first step simulations were conducted to determine the key parameters for laser and detector subsystems. After characterisation of the identified components, a prototype system was designed and installed at the LINAC4 3\,MeV testbench.
In comparison to conventional systems for emittance measurement, the proposed laserwire system has major advantages. Because of its principle, space charge perturbations are excluded. Its range of application starts at beam energies of MeV and reaches beyond Multi-GeV. Since no mechanical parts intercept the beam, it is a reliable and fully non-destructive method.\\
The chosen laser system with kilohertz pulse repetition rate allows to to probe the emittance value of the \hmm beam pulse with microsecond resolution. The fibre based laser system increases the reliability of the laser transport system and reduces greatly its complexity.\\
Thanks to its diamond-based detector system with high bandwidth and sensitivity the presented system works with 3 orders of magnitude less laser peak-power than previous laser-based  systems, which marks a rise of the state of the art in this field.\\

The prototype system was successfully commissioned at the LINAC4 3\,MeV beam. A comparison of phase-space and beam profile with the conventional slit and grid technique showed very good agreement. The twiss parameters and normalized emittance showed reasonable agreement taking into account the systematic error sources for this first prototype. \\
The following commissioning stages at 50~MeV and 100~MeV at LINAC4 will be used to test a modified system for beam profile measurements by collecting the detached electrons. Subsequently further improvements for the final system design will be conducted. Main targets will be the detector resolution and performance as well as the range of the fibre-based laser transport. Finally an operational system is planned to be installed in the 160\,MeV area of LINAC4 to monitor constantly the transverse emittance of the \hmm beam at its future operation.

\section{Acknowledgements}
We want to thank the LINAC4 operations team to give us the opportunity to benchmark our system, the FETS collaboration for lending us their laser, Peter Savage and Richard Epsom for support with design and manufacture of final focus assembly, the BI-group at CERN for their engineering support and Benjamin Cheymol and Francesca Zocca for the slit and grid reference measurements. Moreover we acknowledge the support of the Marie Curie Networks LA3NET and oPAC which are funded by the European Commission under Grant Agreement Number 289191 and 289485.\\
Especially we would like to acknowledge the contribution of Christoph Gabor to this work and the development of laserwire technologies generally, who sadly passed away before this paper could be published. 
\bibliography{laser_bib}
\end{document}